\documentclass[namedreferences]{kluwer}
\newdisplay{guess}{Conjecture}
\usepackage{amssymb}

\newif\ifAMStwofonts
\AMStwofontstrue

\def\ee #1 {\times 10^{#1}}
\def\ut #1 #2 { \, \mathrm{#1}^{#2}}
\def\u #1 { \, \mathrm{#1}}

\def\micron {\, \mu \hbox{m}}

\let\grad=\nabla
\def\cross{\mathbf{\times}}
\def\curl #1 {\grad \cross #1}
\def\div #1 {\grad \cdot #1}
\def\nh{{n_{\rm H}}}

\def\nh{{n_{\rm H}}}

\def\v{\mathbf{v}}

\def\B{\mathbf{B}}
\def\E{\mathbf{E}}            

\def\Bh{\mathbf{\hat{B}}}

\def\Epa{\mathbf{E'_\parallel}}  
\def\Epe{\mathbf{E'_\perp}}  
\def\J{\mathbf{J}}

\newcommand{\delt} [1] {\frac{\partial #1}{\partial t}}

\newcommand{\la} {\lsim}
\newcommand{\ga} {\gsim}

\usepackage{graphicx}
\begin{document}
\begin{article}
\begin{opening}
\title{Star Formation and the Hall Effect}
\author{Mark \surname{Wardle}}
\institute{Physics Department, Macquarie University, NSW 2109, Australia}

\runningauthor{Mark Wardle}
\runningtitle{The Hall Effect in Star Formation}

\date{2003 Jun 20}
\begin{abstract}
The breakdown of flux-freezing in molecular clouds and protostellar 
discs is usually approximated by ambipolar diffusion at low densities 
or by resistive diffusion at high densities.  Here I discuss an 
intermediate regime in which the Hall term in the conductivity tensor 
is significant, and the vector evolution of the magnetic field -- and 
therefore the evolution of the system under consideration -- is 
dramatically altered.  Calculations of charged particle abundances in 
dense gas in molecular clouds and protostellar discs demonstrate that 
Hall diffusion is important over a surprisingly broad range of 
conditions.
\end{abstract}


\end{opening}

\section{Introduction}

The magnetic field in molecular clouds provides pressure support 
against gravity and carries away angular momentum prior to and during 
the collapse of cloud cores to form stars.  Magnetic fields may also 
drive the evolution of protostellar discs through the 
magnetorotational instability, dynamo activity, or by magnetic 
launching and collimation of jets from their surfaces or inner edges.

The diffusion of magnetic field through the weakly ionised molecular 
gas plays a crucial role, allowing gas to slip through the magnetic 
field, or vice-versa.  It is usually considered using one of two 
limits: (i) ambipolar diffusion, in which the magnetic field is frozen 
into the charged species and drifts along with them through the 
neutrals; and (ii) resistive diffusion, in which charged particles are 
completely decoupled from the field by collisions with neutrals.  Both 
limits neglect Hall diffusion, which qualitatively changes the 
evolution of the magnetic field from any given initial 
configuration and has profound implications for the 
magnetically-mediated processes associated with star formation.

Here I present some calculations illustrating the relevance of Hall 
diffusion to molecular clouds and protostellar discs.

\section{Hall Diffusion}
\begin{figure} 
\centerline{\includegraphics[width=10cm]{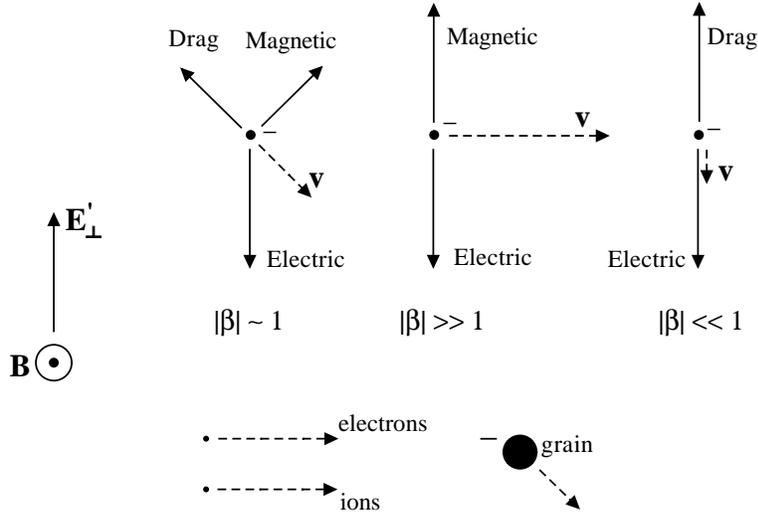}}\vskip 3mm 
\caption{Charged particle drifts in a weakly ionised gas.  The upper 
panel shows the drift speed and force balance for negatively charged 
particles with different hall parameters (see text).  At typical 
molecular cloud densities ions and electron drifts are set by a 
balance between the electric and magnetic forces and they drift 
perpendicular to both the electric and magnetic fields (lower panel).  
The grain drift is affected by collisions with the neurtral particles, 
the resultant drag causes them to drift at an oblique angle to $\Epe$}
\label{fig:hall}
\end{figure}

The diffusion of a magnetic field in weakly ionised gas is determined 
by the drift of charged particles through the dominant neutral 
component in response to the electric field in the neutral rest frame, 
$\E'$.  The drift speed parallel to the magnetic field is set by the 
drag associated with neutral collisions.  In the plane perpendicular 
to the magnetic field, the electric force $Ze\E'_{\perp}$ on a 
particle of charge $Ze$ and mass $m$ is balanced by the magnetic and 
drag forces (see Fig.\ \ref{fig:hall}).  The relative importance of 
these two forces is determined by the Hall parameter
\begin{equation}
    \beta = \frac{ZeB}{mc}\frac{1}{\gamma\rho}\,.
    \label{eq:Hall}
\end{equation}
In the limit $|\beta| \gg 1$, the drag force is negligible, and the 
drift speed $\v$ satisfies $\E_{\perp}'+\v\mathbf{\times}\B/c = 0$.  In 
this case the charged particle is tied to the magnetic field line.  
In the other limit $|\beta|\ll 1$, the magnetic force is negligible 
and  the drag force $\gamma m\rho\v = Ze\E'_\perp$.

At typical molecular cloud densities, ions and electrons have 
$|\beta|\gg 1$ and drift perpendicular to both $\B$ and $\E'$.  
Charged grains, however, have a large collisional cross-section, so 
that drag is much more important and $|\beta|\la 1$.  Grains therefore 
drift obliquely to $\Epe$.  The grain population acquires a 
significant charge through the sticking of electrons, and the ensemble 
of drifting grains modifies the vector relationship between the 
current density $\J$ and $\E'$.  This relationship can be expressed in 
terms of a tensor conductivity $\J = \sigma \E'$:
\begin{equation}
	\J = \mathbf{\sigma}\cdot \E' = \sigma_{\parallel} \Epa + 
	\sigma_\mathrm{H} \Bh \cross \Epe + \sigma_\mathrm{P} \Epe  \,,
	\label{eq:J-E}
\end{equation}
where $\Epa$ and $\Epe$ are the components of $\E'$ parallel and
perpendicular to $\mathbf{B}$, and $\sigma_{\parallel}$, 
$\sigma_\mathrm{H}$ and $\sigma_\mathrm{P}$ are the field-parallel, Hall, and 
Pedersen conductivities, respectively.

The magnetic field evolves according to
\begin{equation}
\delt{\B} = \curl (\v \cross \B) - c \curl \E' \,.
	\label{eq:induction}
\end{equation}
The departure from flux-freezing -- magnetic diffusion -- arises from
the second term on the right hand side and is determined by the
magnetic field configuration via the conductivity tensor: $\E'
= \sigma^{-1}\curl\B/4\pi$.  \ There are three distinct diffusion
regimes:
\begin{enumerate}
    \item  ambipolar diffusion:
	 $|\sigma_\mathrm{H}| \ll \sigma_\mathrm{P} \ll \sigma_\parallel$;

    \item  hall diffusion:
	 $\sigma_\mathrm{P} \la |\sigma_\mathrm{H}| \ll \sigma_\parallel$; and

    \item  resistive diffusion:
	 $|\sigma_\mathrm{H}| \ll \sigma_\mathrm{P} \approx \sigma_\parallel$\,.
\end{enumerate}
The vector evolution of the magnetic field in the presence of Hall 
diffusion is quite different from the other two regimes.  Many of the 
magnetic field geometries adopted in studies of ambipolar or resistive 
diffusion -- in which the magnetic field lies within a plane of 
symmetry --
break down when Hall drifts are present.The evolution is not invariant 
under global reversal of the magnetic field and plane-polarised damped 
Alfv\'en waves do not exist (Wardle \& Ng 1999).  The consequences of 
Hall diffusion during star formation and the subsequent evolution of 
protostellar discs are profound, as found for example in the 
magnetorotational instability  (Wardle 1999; Balbus 
\& Terquem 2001; Salmeron \& Wardle this volume).

\section{Charged species and conductivity in molecular clouds}

\begin{figure*}
\centerline{\includegraphics[width=12cm]{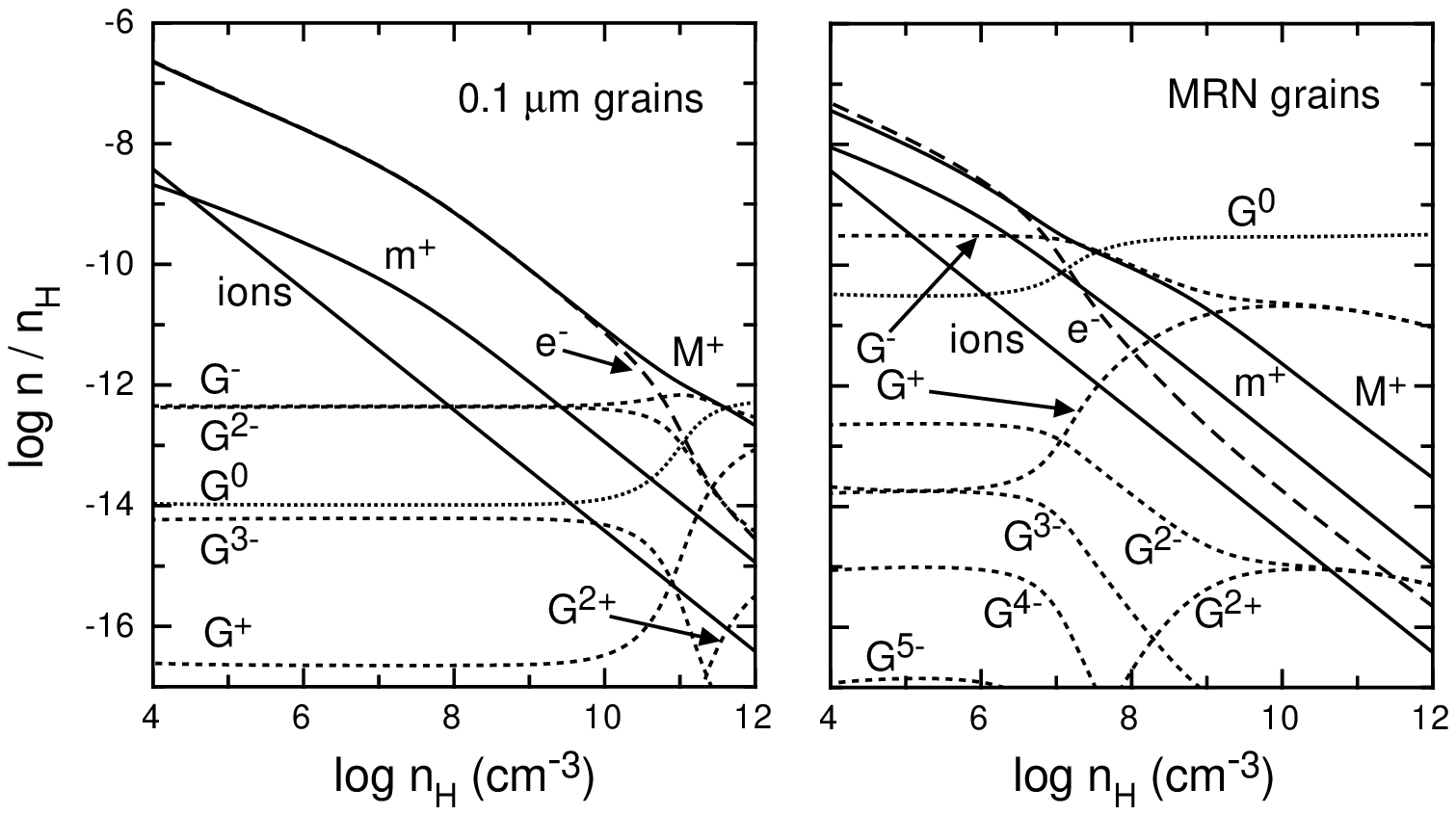}}\vskip 0cm 
\caption{Abundances of charged species in molecular clouds for 
(\emph{left}) 0.1 $\micron$ and (\emph{right}) MRN grain models.}
\label{fig:ionisation}
\end{figure*}
The degree of Hall diffusion in molecular clouds depends on charged 
particle abundances; especially of grains.  Fig.\ \ref{fig:ionisation} 
shows the abundances of charged species in molecular clouds for two 
standard grain models: 0.1$\micron$ grains, and grains 
with an MRN size distribution.  These ionisation equilibrium 
calculations follow Umebayashi \& Nakano (1990) and Nishi, Nakano \& 
Umebayashi (1991) in following the abundance of H$^+$, H$_3^+$, 
He$^+$, C$^+$, representative molecular (m$^+$) and metal (M$^+$) 
ions, electrons, and grains.  The calculations have been improved by 
allowing for the higher grain charge states that appear at 
temperatures in excess of 10 K. Here I have adopted an ionisation rate 
$\zeta = 10^{-17}\ut s -1 \ut H -1 $ and gas temperature 30 K. At this 
temperature, 0.1 $\micron$ grains acquire one or two electrons via 
sticking of electrons.  At low densities this has negligible effect on 
the gas phase ionisation levels because of the relatively small 
numbers of grains ($n_{g}/n_\mathrm{H} \sim 10^{-12}$).  Once the 
density exceeds $10^{11}\ut cm -3 $, the sticking of electrons 
and ions to grains becoming increasingly important, with 
recombinations occurring predominantly on grain surfaces rather than 
in the gas phase.  An MRN grain size distribution contains many more 
small grains which typically acquire a single electron.  These more 
numerous grains dominate the recombination process at lower densities 
($n_\mathrm{H} \ga 10^7 \ut cm -3 $).

\begin{figure*}
\centerline{\includegraphics[width=10cm]{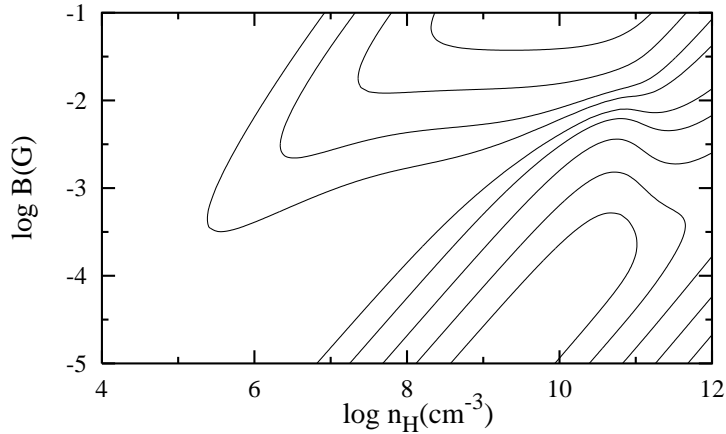}}\vskip 0cm
\caption{Ratio of the Hall and Pedersen
conductivities for the 0.1 $\mu$m grain model.  The contour levels
are set at $|\sigma_\mathrm{H}|/\sigma_\mathrm{P} = $0.01, 0.1, 1, 10
and 100.}
\label{fig:UNplane}
\end{figure*}

\begin{figure*}
\centerline{\includegraphics[width=10cm]{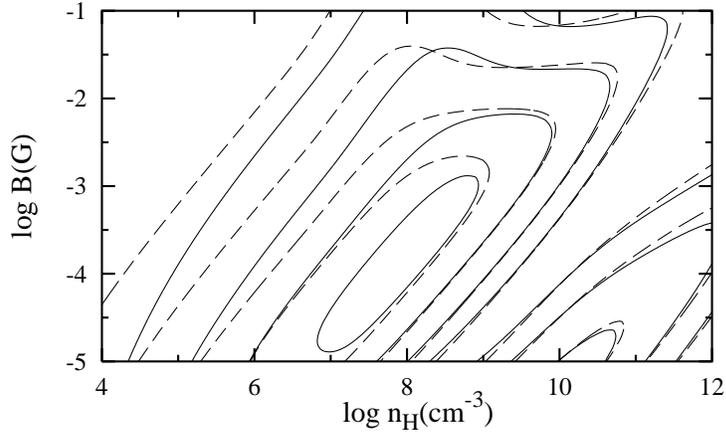}}\vskip 0cm 
\caption{Ratio of the magnitudes of the Hall and Pedersen 
conductivities for an MRN grain model (\emph{solid contours}) and for 
MRN grain distribution with ice mantles (\emph{dashed contours}).  The 
contour levels are set at $|\sigma_\mathrm{H}|/\sigma_\mathrm{P} 
=$0.01, 0.1, 1 and 10.  }
\label{fig:MRNplane}
\end{figure*}

The conductivity tensor for the two grain models can be calculated 
once the magnetic field is specified.  It's instructive to consider 
the ratio of the Hall and Pedersen conductivities ($\sigma_\mathrm{H}$ 
and $\sigma_\mathrm{P}$ as a measure of the significance of the Hall 
effect in magnetic diffusion.  Contours of 
$|\sigma_\mathrm{H}|/\sigma_\mathrm{P}$ are plotted in the 
$B-n_\mathrm{H}$ plane for the 0.1 $\micron$ and MRN grain models in 
Figs.\ \ref{fig:UNplane} and \ref{fig:MRNplane} respectively.  In 
light of the qualitative changes to magnetic field evolution, I 
recommend feeling nervous about neglecting Hall diffusion when 
$|\sigma_\mathrm{H}|/\sigma_\mathrm{P}$ is above 0.1, the second contour 
level in these plots.

For the 0.1 $\micron$ grain model (Fig.\ \ref{fig:UNplane}), two
islands of Hall diffusion are apparent.  The left-hand island follows the
locus where the grain Hall parameter is of order unity and grains
partially decouple from the magnetic field.  The island is truncated
at low densities where the relative abundance of charged grains
becomes too small to influence the conductivity significantly (see Fig. 
\ref{fig:ionisation}).  The right-hand island occurs where the ion
Hall parameter is of order unity.  A zero-line occurs between these
two islands where the contributions of grains and ions to
$\sigma_\mathrm{H}$ cancel because they carry charge of opposite
signs.

The location and size of the grain island depends on the grain size 
distribution, as size determines both the typical charge 
acquired by sticking of thermal electrons and ions and the drag 
coefficient with the neutrals.  In Fig.\ \ref{fig:MRNplane}, the solid 
contours show $|\sigma_H|/\sigma_P$ for an MRN size distribution.  
This has many small grains that acquire a single negative charge (see 
Fig.  \ref{fig:ionisation}), so the grain island extends to lower 
densities but falls off at the highest densities when the abundances 
of positive and negatively charged grains are almost equal and their 
Hall contributions cancel.  The addition of ice mantles to the grains 
(dashed contours in Fig.\ \ref{fig:MRNplane}) tends to extend the 
island to the left, as grains have a higher drag coefficient and 
decouple at lower densities for a given magnetic field strength.

Clearly, Hall diffusion is significant for both the 0.1 $\micron$ and
MRN grain models at densities $\ga 10^6 \ut cm -3 $ and the 0.1-10 mG
magnetic field strengths in molecular clouds.  However, the effect is
reduced if there are many more small grains (such as PAHs, or an MRN
distribution that extends down to a few Angstrom), or if the
characteristic grain size is much larger than $0.1\micron$.  In the
former case, small grains are the dominant charged species at all
densities (see Nishi, Nakano \& Umebayashi 1991) and the near-equality
in abundances of positively- and negatively-charged grains supresses
the Hall effect.  In the latter case, grains are sufficiently
rare that the grain island is shifted to $\ga 0.1\,$G,
in excess of molecular cloud magnetic field strengths.

\section{Conductivity in protostellar discs}
Magnetic coupling in protostellar discs is thought to be poor because 
of two contributing factors.  First, the magnetic field must couple to 
the gas on timescales orders of magnitude shorter than those in 
molecular clouds.  Second, because the gas density is high, 
recombination occurs more rapidly and the fractional ionisation is 
low.  Third, within a few AU of the central protostar the disc column 
density shields the gas from cosmic-rays or x-rays.  This has led to a 
picture of magnetic activity occurring only in the surface layers over 
much of the disc (Gammie 1996; Wardle 1997)

By way of illustration, let us examine the conductivity tensor in a 
minimum-mass solar nebula model at 1 AU from the central star.  The 
temperature and midplane density are $\nh = 6\ee 14 \ut cm -3 $ and 
$T=280\u K $ respectively.  The disc is assumed to be ionised from 
above and below by cosmic rays at a rate $10^{-17} \exp (-\Sigma(z) / 
96 \u g \ut cm -2 )\ut s -1 \ut H -1 $ where $\Sigma(z)$ is the 
surface density from height $z$ to the surface, by x-rays from the 
protostar according to Fig.\ 3 of Igea \& Glassgold (1999) and by 
natural radioactivity at a uniform rate $10^{-21}\ut s -1 \ut H -1 $.

In principle, the grain population is determined by competition 
between growth or sublimation of ice mantles, sticking, shattering, 
gravitational settling to the disc midplane, and stirring by 
convection or turbulence (e.g.\ Weidenschilling \& Cuzzi 1993).  For 
the sake of definiteness, I consider two simple models: (i) 0.1 
$\micron $ grains, and (ii) no grains (assuming that they have settled 
to the midplane).  In the latter case, metal atoms play a key role in 
determining the ionisation fraction in the absence of grains and are 
conservatively assumed to be depleted by a factor of $10^3$ over the 
interstellar value.

\begin{figure*}
\centerline{\includegraphics[width=8cm]{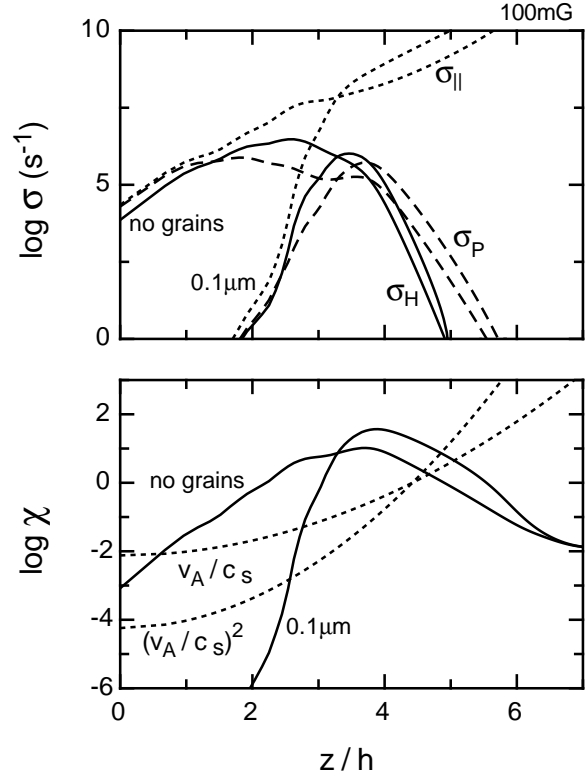}}\vskip 0cm
\caption{ \emph{Upper panel:} conductivity tensor components as a
function of height above the midplane at 1AU in a minimum solar nebula
for a dusty disc with 0.1$\micron$ grains and a model in which grains
have stlled to the midplane.  \emph{Lower panel:} The vertical profile
of the coupling parameter $\chi = \omega_c / \Omega$.}
\label{fig:BHplot}
\end{figure*}
The resulting vertical profile of the conductivity tensor in these two
models is plotted in the top panel of Fig.\ \ref{fig:BHplot} for a
uniform 100\,mG magnetic field.  The much larger conductivities in the
no-grain model within three scale heights of the midplane reflects the
lack of grains -- the charge is carried by more mobile free ions and
electrons.  The decoupling of ions from the magnetic field for
densities $\ga 10^{9}\ut cm -3 $ means that Hall component is larger
than the Pedersen component beween 1.5 and 4 scale heights, and is 
80\%
of $\sigma_\mathrm{P}$ even at the midplane.  In the 0.1 $\micron$
grain model, the Hall conductivity dominates the Pedersen conductivity
within 4 scale heights of the midplane.

One measure of whether the conductivity is sufficient for the magnetic
field to interact with the disc material, is whether the field is
unstable to the magnetorotational instability.  This is determined by
comparing the coupling parameter $\chi = \omega_c /\Omega$ to the ratio
of Alfv\'en speed to sound speed, $v_A/c_s$.  Here $\omega_c$ is the
frequency above which ideal MHD breaks down and $\Omega$ is the
Keplerian frequency.  If Hall diffusion is unimportant the criterion
is $\chi\ga v_A/c_s$, if it is dominant (and the magnetic field has
the correct orientation), then $\chi\ga (v_A/c_s)^2$ is necessary.

The coupling parameter is plotted as a function 
of height in the lower panel of Fig.\ \ref{fig:BHplot}.  The entire 
disc is magnetically active in the settled-grain case, whereas in the 
single-size grain model the layers above 2.5 scale heights are 
active.  In both cases, Hall diffusion is important throughout the 
active regions.

\end{article}
\end{document}